# Public and Teacher Response to Einsteinian Physics in Schools

Alexander Foppoli[1], Rahul Choudhary[1], David Blair[1], Tejinder Kaur[1], John Moschilla[1] and Marjan Zadnik[1]

[1]*The University of Western Australia, 35 Stirling Highway, Crawley, WA 6009, Australia.*

Email: 20930469@student.uwa.edu.au

## Abstract

Einsteinian physics represents a distinct paradigm shift compared to Newtonian physics. There is worldwide interest in introducing Einsteinian physics concepts early in school curriculum and trials have demonstrated that this is feasible. However introducing Einsteinian concepts from an early age requires more than suitable curriculum and teaching resources - it also requires teacher training and public support. This paper describes a pilot study used in an attempt to gauge public and teacher support. This entailed giving teachers, who included the entire staff of a primary school, and self-selected family groups an in-depth understanding of proposed curriculum content through public outreach and professional development workshops. We assessed their attitudes through questionnaires. Comments and opinions from the public were also collected from online resources. Results show overwhelming support from both teachers and the public. We assessed attitudes of children as well as adults and obtained opinions regarding the appropriate age at which to begin to introduce Einsteinian concepts.

Keywords: Einsteinian physics, physics education, models, analogies, attitudes, family learning, public science education

## 1. Introduction

A decline of school student enrolment in Science, Technology, Engineering and Mathematics (STEM) subjects is a cause for alarm in societies dependent on technical skills according to the European Directorate-General for Education (2018) and the U.S. Department of Education (2016). *Australia's National Science Statement* from 2017 reports STEM enrolment for students is the lowest in 20 years (Australian Government, 2017).

Researchers in Australia (Kaur et al, 2017.a & 2017.b), Germany (Zahn and Kraus, 2014), Norway (Bungum et al., 2015 and Henriksen et al., 2014) and the United States (McKagan et al., 2008) are investigating the implementation of various Einsteinian education programs into the classroom.

It is widely suggested that the decline in student STEM involvement can be reversed with the introduction of exciting science programs aimed at maximizing the interactive learning experience as reported by Angell et al. (2004), Lavonen et al.

(2004) and Geelan (1997). The Einstein-First Project aims to teach the basic concepts of modern physics from an early age. The reason for this is a) to teach them the language and concepts of our best understanding of the nature of space, time, matter and radiation, b) to prevent the conceptual conflicts associated with the transition from the commonly taught Newtonian world view, to the radically different Einsteinian world view (which currently only a small minority of students get to learn at senior or tertiary level), c) to make school science relevant to the modern world of mobile phones and other technologies that can only be understood in terms of Einsteinian physics.

The challenge of teaching Einsteinian physics required the development of many teaching resources. We developed a program based on extensive use of models and analogies, and active learning (Kaur et al, 2017.a & 2017.b). The goal of the Einstein-First project is to facilitate widespread implementation of an Einsteinian based curriculum.

Before any educational program can be introduced to students, it must satisfy several requirements. Firstly the program needs to have been fully tested. Secondly the program needs to have teacher support and associated teacher training and thirdly it must have public support. The purpose of this paper is to report an assessment of the latter two requirements, which will be discussed further in section 2.

The Einstein-First Project grew out of gravitational wave research. The discovery of gravitational waves in 2015, one century after Einstein's general theory of relativity and the concepts of black holes and gravitational waves were first published, was a momentous achievement. More than 1000 physicists globally worked together in the LIGO-Virgo collaboration to achieve this detection (Abbott et al., 2016), which was awarded the 2017 Nobel Prize for Physics (Nobelprize.org, 2018). The discoveries in the first two years have directly observed black holes, have confirmed that gravity travels at the speed of light and detected the coalescence of neutron stars (Abbott et al., 2017) in which the spectra of heavy element nucleosynthesis was observed (Kasen et al., 2017). The gravitational wave detectors themselves harness the quantum properties of light to measure the smallest amount of energy ever detected (Harry, 2010). The era of gravitational wave astronomy in which we can listen to gravitational waves from throughout the universe has just begun.

In this new era it becomes even more important that people are equipped with the conceptual framework with which to comprehend these new discoveries. It is somewhat absurd for school curricula to contain content based on absolute space and time, and Euclidean geometry, and gravity as a force that is transmitted instantaneously, when news bulletins report detections of ripples in space, black holes and measurement of the speed of gravity (Weule, 2017). T. Kaur et al (2017.a, 2017.b & 2017.c) and Pitts et al. (2013) have shown that Newtonian gravity can be easily taught within an Einsteinian context using space-time curvature models, while Stannard et al (2016) have shown that gravity and free fall can be understood within the framework of gravitational time dilation. Thus Newtonian and Euclidian concepts can be learnt as useful approximations within an Einsteinian worldview.

Similarly the concept of light as a stream of photons can also be easily integrated into school science as long as it is combined with an appropriate understanding of interference. Without this concept students cannot understand solar panels, digital cameras and many other aspects of modern technology. Early introduction of the photon concept developed by Einstein (1905) can be combined with the atomic description of matter in which all forms of matter and radiation share two fundamental properties: wavelength and momentum that are related by de Broglie's (1970) formula,

$$\text{wavelength} = \text{Planck's constant}/\text{momentum}.$$

Results on student attitudes and learning in trials at various levels from ages 11 to 16 have shown that students easily grasp the fundamental concepts, and that it has positive effects on student attitude (Pitts et al., 2013), as well as gender parity reported by Kaur et al. (2017c) and Choudhary et al. (2018).

This paper reports on a pilot study involving a series of educational workshops presented to the public and schoolteachers. The programs were designed to showcase the Einstein-First program using a subset of the content used in the classroom. In the Einstein-First program students learn through role-playing, models and analogies, discussions, video presentations, lab tours and practical experiments. The showcase program was designed to provide the participants with a deep insight into the Einstein-First program currently being taught to students. The showcase program was designed to be educational in its own right, allowing participants to provide valid responses and opinions about the program.

In section 2 we will further discuss the rationale for testing both public and teacher attitudes to learning Einsteinian physics in terms of the likely requirements for implementing widespread curricula change. Section 3 describes the methodology used to measure the public, teacher and student response to the showcase program. Participant responses are discussed in section 4 along with on-line responses to articles describing the program.

## 2. The requirements of implementing an educational program

### 2.1 Proven programs
As mentioned above, testing of the teaching of Einsteinian physics has yielded positive outcomes for students aged 11 to 16, the Einstein-First program began with initial trials with 11 year old students and results were reported by Pitts et al. (2013) Following this program, the approach, which is based on extensive use of models and analogies, were developed and tested across various levels of middle schools (Kaur et al, 2017.a & 2017.b). While further testing is continuing, it is clear from the work to date that the programs are extremely effective, not only demonstrating high levels of student knowledge and acceptance of core concepts, but also improvement in attitude to science. Positive results, observed by Pitts et al. (2013), Kaur et al. (2017.c) and Choudhary et al. (2018), from these programs have shown an increase in student knowledge and attitudes, which indicates that the program is already quite effective in communicating the ideas of Einsteinian physics.

### 2.2 Teacher response
The successful programs discussed above are not sufficient without ensuring the ability of teachers to cope with a paradigm shift, which includes approaching topics which they may never have encountered and never been asked to teach such as the nature of space, time dilation and the concepts of quantum mechanics. Without being comfortable with the content, teachers will be ineffective in communicating ideas for which the common prejudice holds that they are very difficult. Without experience of the Einstein-First program itself, teachers are likely to be daunted, incorrectly believing that it requires high mathematical skill and exceptional intelligence. It is vital that teachers feel comfortable with the concepts they may be asked to teach to their students. In addition to a teacher's knowledge, we must ensure that they have a positive attitude towards the teaching of these topics.

### 2.3 Public Response
The school curriculum is a response to public demand (Newmann, 1975); members of the wider community can be contributing to the overall wealth of societal knowledge, whilst also simultaneously learning from it. The introduction of new curriculum is often fraught with difficulty, and sometimes arises only through public pressure. Examples include socially contentious issues such as sex education (Stephenson et al., 2008), religious education (Lankford and Wyckoff, 1992) and in biology the introduction of modern genetics (Freidenreich, Duncan and Shea, 2011). Concern over the classical status of physics is manifested in documents such as the "Open Letter to the President" Youtube video (Minutephysics, 2012). Some people, including scientists, take the view that everybody needs to learn classical physics before they are ready to take on the modern paradigm (Ligare, 2016). It is important to know whether this is a widely held point of view. Only if there is broad public support, combined with a positive teacher response can we expect curriculum authorities to begin working towards updating school curricula around the world.

## 3. Methodology

Here we will discuss the structure of the showcase programs, the method of assessment used in this pilot study and our approach to examining the validity of the data. The showcase programs were presented through Public Outreach Workshops (POW) and Professional Development Workshops (PDW) held at the Gravitational Wave Research Facility at Gingin, Western Australia and the University of Western Australia during 2017. The programs in each case were similar.

### 3.1 Program Outline
The POW and PDW programs were designed to teach the fundamentals of Einsteinian physics and gravitational wave detection through a combination of presentations and activities. These were designed to teach the following concepts:

1. Light comes as photons.
2. Photons have momentum.
3. Interference patterns occur when light can take two or more paths.
4. Space-time is a four dimensional fabric in which all forms of energy and matter exist.
5. Gravity is caused by mass curving space-time.
6. Gravitational waves are ripples of space-time, these can be detected with light interference.

These ideas are taught through the activities mentioned below, some of the material is also available online (The Einstein-First Project, 2018) and further described in other papers by Kaur (2017.a &2017.b) and Choudhary (2018).

### 3.1.1 Video/interactive presentation

Participants were introduced to the concepts of general relativity and quantum mechanics through a combination of educational videos, texts and images. Presenters were used to further describe the phenomena being displayed through the visual aids, participants were permitted to ask questions and explore these theories further through group discussions.

### 3.1.2 Role-Play

Self-selected participants wore simple costume props (such as an Einstein wig) and recited a script to the other workshop participants. Participants re-enacted Heinrich Hertz's discovery of electromagnetic waves and then presented Albert Einstein's explanation of how the quantization of light could be used to describe Hertz's observation of the photoelectric effect. The role-play also featured Richard Feynman who was a proponent of a particle understanding of light and proved the theoretical reality of gravitational waves, which were also predicted by Einstein.

### 3.1.3 Model Interferometer

Participants used a laser pointer, plastic beam splitter and two Styrofoam hanging mirrors with the aim of directing the light to hit both mirrors at the same time. This model interferometer introduced the concepts of gravitational wave detection and shows the fundamentals of how a Michelson interferometer is used to detect ripples in space-time. This activity also allowed users to fire a Nerf gun at a single hanging mirror and observe the movement of the mirror; this taught the concept of photons having momentum and the ability to push objects.

### 3.1.4 Light Experiments

A set of experiment activities focussed on interference created with very simple apparatus. They are designed to explore the particle and wave-like nature of light and also to show how wave-like properties can be used to make precise measurements. Participants observe interference patterns created with laser pointers. They calculate the width of their hair from the observed interference and make captivating interference patterns by reflecting laser beams off soap films. The fact that interference occurs even with single photons is used to emphasise the counter-intuitive aspects of quantum physics.

### 3.1.5 Space-time Simulator

The concepts of Einstein's theory of gravity are explored using a stretch lycra sheet space-time simulator. It creates a visual representation of the curvature of space-time in the presence of mass. Using a variety of steel balls, the simulator allows participants to visualize the fundamental connection between matter and space-time curvature. Participants used balls on the lycra sheet to observe geometrical warping and numerous orbital experiments, including orbital patterns, geodetic precession and tidal locking, all of which are naturally observed in the universe.

**3.2 Public Outreach Workshop**
The POW involved a self-selected sample group of the public that was invited to attend through online and newspaper marketing posters. The program took place at the Gravitational Wave Research Facility in Gingin, Western Australia and ran for approximately two hours. The programs ran recurrently over two days, from which 20 students and 25 adults that attended the workshop completed attitudinal questionnaires.

The POW included the activities listed above (3.1.1-3.1.5) and additionally involved:

   3.2.1 Spatially Linked Learning (SLL)
   This activity used a series of signs at the beginning of the program designed to introduce the readers to the facility and teach them about the fundamentals of gravitational waves. 25 signs were separated by the approximate time it takes to walk to the next sign, this aimed to give the reader enough time to understand the content of each sign. While we did not retain quantitative data, we did observe young people, in particular, running from sign to sign and discussing their content. The ordered placement of the signs was used to direct the reader to the entrance of the Gravitational Wave Research Facility and also to link the reader's memory to their spatial awareness (Kelly, 2016). Examples of the text on the signage included:

   *"Gravitational waves were predicted by Albert Einstein in 1916"*
   *"Gravitational waves are ripples of space-time"*
   *"The instrument used to detect gravitational waves is called an interferometer"*

   3.2.2 External Lab Tour
   This activity involved an external tour of the Gravitational Wave Research Facility; this tour explored the application of Einsteinian physics into experimental research. The tour began with a model of vibration isolation technology, which explored the physics involved in suppressing seismic vibration. The participants then visited Kip Thorne's tree, which was used for the inauguration of the Research Facility. They were shown an educational plaque that explains how accelerating masses generate gravitational waves. Participants were then shown the long vacuum pipes along which high-powered lasers are targeted to reflect off test masses to make precision measurements between them.

**3.3 Professional Development Workshop**
The PDW involved two separate programs; both took place at the University of Western Australia. In the first PDW, 9 secondary school science teachers were recruited through a University of Western Australia Outreach Program. A selection of 25 primary schoolteachers from Rosalie Primary School in Perth, Western Australia was invited for the second program.

The Professional Development Workshop (PDW) program included all five of the activities described above (3.1.1-3.1.5.) It also included a discussion of results from research interventions summarized below.

We presented results from Pitts et al. (2013) that demonstrated how year 6 students developed a substantial understanding of time, space, light and gravity. The results showed the majority of students did not think they were too young to understand the Einsteinian concepts. We also discussed results observed by Kaur et al. (2017.c) that demonstrated substantial understanding developed by year 9 students, and by Choudhary et al. (2018) who compared one day interventions across years 7, 8, 9 and 10. All results showed substantial improvement factors for conceptual understanding. Gender analysis showed that females had greater improvement factors than male students, with lower initial scores and near parity with males at the end of the programs. Following this summary, we discussed with teachers how the activities presented could be used in their classrooms.

**3.4 Assessment Process**
All participants in the Public Outreach Workshop and Professional Development Workshop programs were introduced to the Einstein-First curricula through the showcase programs and given questionnaires to complete. The Public Outreach Workshop was assessed using only post-questionnaires because it was considered inappropriate to ask members of the public to complete questionnaires before joining the workshop. Separate questionnaires were designed for students and for adults. Participants of the Professional Development Workshop were asked to complete conceptual understanding pre-questionnaires at the start of the program. At the end of the program, they were asked to complete the same conceptual understanding questionnaire and an attitudinal questionnaire.

The POW attitudinal questionnaires used a Likert-type scale to respond to statements about physics and physics education. The teacher attitudinal questionnaire included questions that required short written responses. Conceptual questionnaires required short written answers to questions on Einsteinian concepts. Further details are given below.

    **3.4.1 Student attitudinal questionnaire**
    The student attitudinal questionnaire included seven statements. These statements were designed to assess student response towards the program and the future teaching of Einsteinian physics in their classroom, such as "*I would like to learn the modern ideas of gravity and light at school*" and "*I think I am too young to understand Einstein's ideas*".

    **3.4.2 Adult attitudinal questionnaire**
    This questionnaire, consisting of twelve statements, was designed to assess the adult public response to the program. The statements explored their attitudes towards science and also whether students should be taught Einsteinian physics at school. Statements on the questionnaire included "*I think everyone should have the chance to know about gravity and light*" and "*studying gravitational waves might make useful discoveries for daily life*".

    **3.4.3 Teacher conceptual pre-questionnaire**
    The conceptual pre-questionnaire was designed to assess the teacher's prior knowledge of Einsteinian physics before they were introduced to the program. This questionnaire was modeled from student questionnaires used in previous Einstein-First programs by Pitts et al. (2013), Kaur et al. (2017.c) and

Choudhary et al. (2018) and asked questions that focused on the fundamental understanding of the universe such as "*what is light?*" and "*what is gravity?*"

### 3.4.4 Teacher conceptual post-questionnaire
The conceptual post-questionnaire had identical questions as the conceptual pre-questionnaire and was designed to assess any change in the teacher's understanding of physics after being involved in the PDW program.

### 3.4.5 Teacher attitudinal post-questionnaire
The attitudinal post-questionnaire was designed to assess the attitudes of teachers towards the Einstein-First program being introduced into the classroom. Examples of some of the questions asked include "*Do you think it is appropriate to teach these modern ideas of Einsteinian physics to students at a young age?*" and "*Would you like to have extra training for teaching these modern ideas?*"

## 3.3 Data Analysis and Validity
The questionnaires we designed were modified from previous student questionnaires in order to be more applicable to the public and teacher audience. Expert researchers reviewed the quality of the questions and statements. The purpose of this was to ensure that the wording of the questionnaires covered the fundamental core conceptual understanding and attitudes that we were trying to investigate. We wanted to ensure that the respondents were able to clearly communicate their opinions through their responses.

The data was processed through Excel to calculate the mean response and the distribution of varying attitudes towards the program. The responses given in the conceptual questionnaires were compared to responses expected from the expert research panel in regards to the fundamental aspects of modern physics. We awarded correct responses to answers that described light as being "*a stream of photons, that travel as particles but also behave like waves*" and described gravity as being "*a curvature of space-time caused by mass*". Completed questionnaires from both workshops were collected; results have been presented and discussed in section 4. Section 4 also contains an analysis of online support provided by public users.

## 4. Results and Discussion

### 4.1 Public Outreach Workshop

### 4.1.1. Adult attitudinal response
As discussed in the introduction, we used the context of gravitational wave detection, including concepts of curved space and quantum measurement, as the vehicle for introducing Einsteinian physics. The responses from 25 parents and members of the public that attended the Public Outreach Workshop are presented below, followed by a discussion of the results.

| Statements | Strongly Disagree | Somewhat Disagree | Neutral | Somewhat Agree | Strongly Agree |
|---|---|---|---|---|---|
| (a) This program gave me an insight into physics research | 0% | 0% | 0% | 24% | 76% |
| (b) This program changed my ideas about gravity | 4% | 0% | 20% | 36% | 40% |
| (c) This program changed my ideas about light | 4% | 0% | 16% | 44% | 36% |
| (d) This program changed my ideas about the universe | 4% | 0% | 24% | 24% | 48% |
| (e) It was interesting to hear about gravitational waves and the nature of light | 0% | 0% | 4% | 8% | 88% |
| (f) I like finding out about physics | 0% | 0% | 0% | 16% | 84% |
| (g) I think it is useful to know about science | 0% | 0% | 0% | 0% | 100% |
| (h) Studying gravitational waves might make useful discoveries for daily life | 0% | 0% | 4% | 16% | 80% |
| (i) I enjoyed doing the experiments | 0% | 0% | 0% | 20% | 80% |
| (j) I think everyone should have the chance to know about gravity and light | 0% | 0% | 0% | 0% | 100% |
| (k) Modern ideas of gravity and light should be in high school textbooks | 0% | 0% | 0% | 8% | 92% |
| (l) I will tell my friends or family about what I learnt today | 0% | 0% | 0% | 20% | 80% |

Table 1 – Attitudinal responses from adults show support for introducing concepts of Einsteinian physics into the school curriculum

The first statement (1(a)) was designed to determine whether the public obtained the insights we were trying to provide. The results are clearly affirmative with 100% of participants either agreeing or strongly agreeing. The next three statements (1(b), 1(c) and 1(d)) were trying to assess both the novelty and the comprehension of the material presented. While there was very strong agreement (76%, 80% and 72%), roughly one quarter of respondents were neutral or disagreed. We take this to be an indication of the fraction of self-selected participants who came with substantial knowledge. Near universal agreement to statement 1(e) shows that there was substantial novelty in the material presented, even for those who had some prior knowledge, about the nature of gravity and light.

The next two statements ((f) and (g)) tested attitudes to physics and science in general. There was universal agreement from respondents in both finding physics interesting and understanding the importance of science, which shows strong support for implementing physics education programs. The next statement 1(h) was designed to investigate public response towards the recent detection of gravitational waves. A near unanimous response (96%) shows that a vast majority of the self-selected participants are excited and that future breakthroughs could be discovered from the gravitational wave technology discussed throughout the program.

The following statement 1(i) was used to explore the adult response to the activities that we have tested with students in previous programs. The universal agreement from respondents suggests strong enthusiasm from the participants towards activity-based learning material and the implementation of this style of learning in our programs.

The next two statements (1(j) and 1(k)) investigate the public attitudes towards the education outreach of Einsteinian physics and the changing of the curriculum. Universally positive responses to both statements suggest that all members of the self-selected group strongly support teaching modern physics to a wider audience and would also like these concepts to become a part of the school curriculum. Statement 1(l) aimed to capture the degree of enthusiasm, as measured by whether they were sufficiently excited by the program to tell their friends and family about it. Again, we see universal agreement. This also indicates a high degree of comprehension because people are unlikely to communicate ideas they do not understand.

As already discussed, the above results are clearly influenced by the fact that the audience was self-selected. Not withstanding this difficulty it is clear that there is negligible disagreement with the concept of teaching Einsteinian physics in school, while the self-selected audience is also strongly of the opinion that societal benefits may occur from the fundamental studies of gravitational waves. Results from adults provide strong evidence in support of implementing the Einstein-First program into schools. The collected responses from self-selected members of the public suggest that the public strongly encourages teaching Einsteinian physics to students. The public attitude towards physics research and its applications appear to have been effectively communicated to the public during the 2-hour workshop.

### 4.1.2 Student attitudinal response

The attitudinal responses from school students to the post-program questionnaire are presented below. (N=20) The results are comparable to the adult responses. Of particular interest is student perception of ability to understand, as discussed further below.

| Statements | Strongly Disagree | Somewhat Disagree | Neutral | Somewhat Agree | Strongly Agree |
|---|---|---|---|---|---|
| (a) It was interesting to find out about gravity and light | 0% | 0% | 5% | 40% | 55% |
| (b) I enjoy doing science experiments | 0% | 0% | 5% | 10% | 85% |
| (c) I think everyone should learn about gravity and light | 0% | 5% | 10% | 20% | 65% |
| (d) I would like to learn the modern ideas of gravity and light at school | 0% | 0% | 0% | 30% | 70% |
| (e) I wish we could do experiments like this at school | 0% | 0% | 0% | 0% | 100% |
| (f) I want to explain about light and gravity to my friends | 10% | 5% | 40% | 10% | 35% |
| (g) I think I am too young to understand Einstein's ideas | 35% | 20% | 5% | 10% | 30% |

Table 2 - Attitudinal responses from students show support for introducing concepts of Einsteinian physics into the school curriculum

The first statement 2(a) was designed to measure student interest in finding out about the concepts of light and gravity. We see near universal agreement, however 40% agreed rather than strongly agreed. Statement 2(b) tested student attitude to science experiments. There was strong agreement with 85% strongly agreeing.
The next three statements (2(c), 2(d) and 2(e)) relate to the curriculum and show that students strongly agree that the material should be in the curriculum. In addition, the experiments they undertook were clearly appreciated, and there is a universal wish that they could do practical experiments like this at school.

The next statement (2(f)) explores the student response towards sharing the ideas presented in the program to their friends. A mixed response suggests that students aren't opposed to discussing science with their friends. The mixed result could also be due to the students not feeling as though they developed a strong enough understanding of these concepts during an introductory program, compared to if they were taught these ideas in their classroom.

Statement 2(g) tested young people's views on ability to understand modern physics at their age. It was deliberately asked in a negative context. Overall, 55% of students disagreed that they were too young.

| Age analysis of statement 2(g) | | | | | |
|---|---|---|---|---|---|
| | Strongly Disagree | Somewhat Disagree | Neutral | Somewhat Agree | Strongly Agree |
| 5-9 year olds | 44% | 0% | 0% | 11% | 45% |
| 10-14 year olds | 11% | 45% | 11% | 11% | 22% |

Table 3 – Age dependence of student attitude to learning Einsteinian physics

Given that 40% of students agreed with statement 2(g), we further analysed the age dependence of responses to this statement as given in Table 3. A small majority of the younger group (56%) agrees that they are too young to understand Einsteinian physics. In the older age group, 56% do not believe they are too young whilst only 33% agree with the statement that they are too young. It is not surprising that a short duration exposure gave mixed results, as the POW provided only a brief and informal introduction.

Overall, the above results show strong support from students towards learning the modern concepts of light and gravity. The student responses also show strong support for activity-based learning and suggest that students would like the school curriculum to include Einsteinian concepts.

## 4.2 Professional Development Workshop

### 4.2.1 Teacher attitudinal response

Following the Professional Development Workshop we tested teachers attitudes to the need and possibility of teaching Einsteinian physics at an early age. The teachers involved in both development programs taught classes between the range of early childhood (5 years old) to grade ten (15 years old). Due to the consistency between attitudes and opinions observed over both PDW programs, the responses for this section have been combined. The teachers were asked four questions as listed below.

| Questions | Yes | No | Unsure | No Response |
|---|---|---|---|---|
| (a) Do you think it is appropriate to teach these modern ideas of Einsteinian physics to students at a young age? | 94% | 3% | 0% | 3% |
| (b) Would you like to use the activity materials for teaching your students? | 94% | 0% | 0% | 6% |
| (c) Would you like to have extra training for teaching of these modern ideas? | 88% | 6% | 0% | 6% |
| (d) Do you think your school would have enough funds to purchase the materials you saw today (say $1000)? | 62% | 6% | 29% | 3% |

Table 4 – Responses from teachers show a positive response towards introducing Einsteinian physics into the classroom, provided that the teaching material is low-cost

The response to question 4(a) show that a large majority (94%) of the teachers supported the concept of Einsteinian physics at primary and middle school. In addition to the responses to question 4(a) as shown in Table 4, some respondents also chose to include a desired year level at which they believe we can begin teaching these concepts. Question 4(a) did not specifically ask respondents to include this, however these responses have been analysed below in Table 5.

| Suggested age to introduce Einsteinian physics | | | | | |
|---|---|---|---|---|---|
| No response | Grade 3 | Grade 4 | Grade 5 | Upper primary | Too young |
| 64% | 9% | 12% | 3% | 9% | 3% |

Table 5 – Teacher responses of the youngest age at which to teach Einsteinian physics

Although the majority (64%) of teachers did not suggest a recommended year level to begin teaching Einsteinian physics in their response as seen above, we can see that a small sample of respondents support teaching these concepts to students as young as Grade 3 (8 years old). This indicates that teachers are not opposed to implementing the Einstein-First program in primary schools.

The strong positive response (94%) to 4(b) indicates teacher enthusiasm for the active teaching approach used in the Einstein-First program. However, as demonstrated by question 4(c), the majority of teachers (88%) believed that they would require additional training and resources in order to feel sufficiently prepared to teach Einsteinian physics in the classroom. We also wanted to determine whether the cost of resources would be a significant impediment. Question 4(d) shows that a majority

(62%) consider the modest costs of equipment not to be a severe impediment, although some teachers were unsure.

Overall the teacher responses were very positive. The large majority of teachers were both enthusiastic about the concepts and the activity based learning approach. Most teachers believed they needed additional training, while many believed their schools could fund the rather low cost equipment required. These results also suggest that some teachers support teaching these concepts to children as young as 8 years old.

**4.2.2 Teacher conceptual understanding response**
We were interested to measure teacher acceptance of the two core concepts, light and gravity, that we presented in the program. To investigate their previous understanding and the capability of the program to explain these concepts, we asked them 1) "*What is light?*" and 2) "*What is gravity?*" both before and after the PDW. The results from the 34 primary and secondary school teachers are presented below.

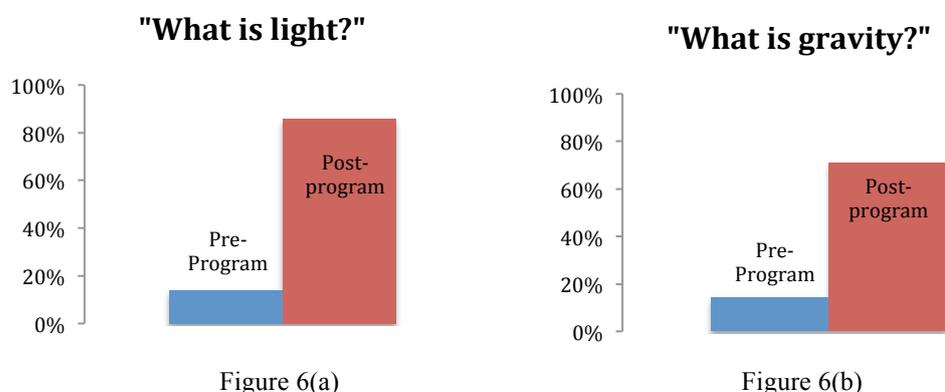

Figure 6(a)　　　　　　　　　　　Figure 6(b)
Figure 6 – The conceptual knowledge of light and gravity for teachers before and after an introduction to the Einstein-First program

Figure 6(a) and 6(b) show that the teachers entered the program with unsophisticated concepts of light (14%) and gravity (14%). Three example answers given by respondents to the question "*What is light?*" were 1) "*A ray*" 2) "*What we get from from the sun*" and 3) "*Not sure how to explain. Opposite of dark*". Two typical answers given to the question "*What is gravity?*" were 1) "*A force that pulls objects to Earth*" and 2) "*It has something to do with the moon.*"

After the program, teachers were able to give a modern answer to both questions, achieving average scores of 86% and 71% respectively. Typical answers to the question "What is light?" after the program were "Photons, travelling like bullets" and "Particles that move and combine to create patterns." For gravity, typical answers were "n*atural motion through curved space*" and "*matter tells space how to curve, space tells matter how to move.*"

The results of the PDW program show that the concepts presented were not intrinsically difficult. There was a very high uptake of the concepts of light and gravity. The pre-test showed that teachers had very low understanding of the modern ideas of both of these concepts. This result is consistent with schools programs that show that school students understanding of the modern concepts is not related to their pre-knowledge. The acceptance of Einsteinian concepts is not related to academic ability, it merely requires exposure to the key concepts.

## 4.4 Online opinions

As this Einstein-First program has been developed, the team has tried to communicate the concepts to a broader base of readers through articles on popular websites. Websites such as the Conversation (2010) provide a powerful tool for analysing opinions. The website has strict guidelines regarding authenticity of material presented and provides convenient means for readers to give responses. One difficulty is that some responses lead to extensive debates between a few vocal and opinionated readers. We have analysed only direct responses to two articles 1) *"Testing the theory: taking Einstein to primary schools"* (Blair, 2012) and 2) *"Why don't we teach Einstein's theories in school?"* (Blair, Henriksen and Hendry, 2016) published on The Conversation. The first article explored the motivation for changing school curriculum to include modern physics. It has been viewed 5,700 times with 22 comments. The second article presents a more international perspective and results obtained from successful implementations of the Einstein-First program. It is co-authored by project leaders from Norway, Scotland and Australia. It had been viewed 18,700 times with 27 comments at the time of analysis.

For analysis we divided comments into five categories 1) strong support, 2) qualified support, 3) too complex, 4) opposed and 5) not relevant. A summary of the 49 responses is given in Table 7.

| Title of Article | Strong support | Qualified support | Too complex | Opposed | Not relevant |
|---|---|---|---|---|---|
| 1. Testing the theory: taking Einstein to primary schools | 45% | 41% | 0% | 0% | 14% |
| 2. Why don't we teach Einstein's theories in school? | 22% | 37% | 15% | 4% | 22% |

Table 7 – Analysis of online comments about both articles on the Einstein-First program

The responses to Article 1 show support from 86% of the comments. None were opposed, while 14% of comments were irrelevant. In the case of the second article, 19% of recipients were opposed, with most comments indicating confusion between mathematical tools and concepts. A typical response in this category was *"How do you teach it without differential equations?"*. The same comment could also apply to Newtonian mechanics. Clearly the reader did not comprehend the difference between conceptual understanding and the skills required to calculate such things as orbital dynamics. It is important that those advocating to Einsteinian physics be aware of this source of confusion.

The following comments show online public support for the implementation of Einsteinian physics.

a) "Excellent – I think it's great to confront kids with these ideas."

b) "I had to wait for university before relativity was explained but I did get the basics of quantum in high school in the late 1960s."

c) *"What a great initiative. It has an important symbolic value: we are the kind of people, populating the kind of society that takes science seriously and believes that children are capable of learning complicated things."*

The above comments indicate that members of the public think that the concepts of modern physics can successfully be taught to students.

Some of the comments identified two significant issues: a) the need to train teachers and b) the difficulty in changing the curriculum. Two examples are given below.

a) *"The issue clearly isn't the student's capacity to learn. Perhaps it's the teacher's capacity to teach…"*

b) *"Given the discovery of gravitational waves only happened this year, the reason we don't teach Einstein's theories in our schools is because of the exorbitant amount of time it takes to develop and approve curriculum in this country."*

These above comments suggest that the support from teachers and the overall curriculum needs to be guaranteed if students are to be taught modern physics at school.

The following comment indicates recognition of the need to retain Newtonian ideas as both tools and concepts.

*"Yes I love these fundamental ideas about space and time and it is important that we help children learn the truth….But knowledge and models also serve to guide everyday behavior and decision-making and Newtonian concepts have real utility here, like understanding the kinetic energy of your car and how a pulley works."*

The following comment indicates recognition that changing the curriculum is not an easy task, as we discussed in the introduction.

*"So this is a good initiative. However, I think there is a lot more work to do."*

The comments analysed above clearly represent a self-selected audience. Many comments were useful in identifying difficulties. However, overall, the lack of opposition implies strong public support.

## 5. Conclusion

We have evaluated public responses to the idea of teaching the fundamental concepts of Einsteinian physics at school. Opinions were obtained through both exposure to a showcase program and through public responses to expert articles on the Conversation website. Responses show strong support for the teaching of Einsteinian physics. Young people participating in the public outreach interventions also showed extremely strong positive responses.

In a similar pilot study, we evaluated the response of primary and middle school teachers to the program. The teachers were asked whether they could teach the concepts in school. Amongst participating teachers, we found a very positive response, with teachers considering that the concepts could be taught at various levels of primary school, with teachers suggesting various minimum age levels from grade 3 to upper primary.

Overall, the results presented here show that both parents and children support the teaching of modern concepts of gravity and light in school. A large majority of the primary school teachers surveyed who were exposed through our professional development programs think that it is appropriate to teach Einsteinian physics to their students, although 88% would like to have extra training.

This study has shown that the core concepts of modern physics are within reach of teachers across all teaching ages, while analysis of public outreach workshops show strong support for teaching Einsteinian physics at schools. If the broader public was enfranchised by being given a deeper understanding of the concepts that underpin our modern life, through universal education, it might be expected that attitudes to science would improve.

In response to this study, we have gone on to plan a more extensive public outreach study with a much larger size, as well as additional teacher professional development programs to further investigate teacher attitudes. In addition we have run a 3-week intervention with 8-year-old students to test the ability to take up core concepts at the youngest age suggested by teachers.

**References**
Abbott, B., Abbott, R., Abbott, T., Abernathy, et al (2016). Observation of Gravitational Waves from a Binary Black Hole Merger. *Physical Review Letters*, 116(6).
Abbott, B., Abbott, R., Abbott, T., Acernese, et al (2017). GW170817: Observation of Gravitational Waves from a Binary Neutron Star Inspiral. *Physical Review Letters*, 119(16).
Angell, C., Guttersrud,  Henriksen, E. and Isnes, A. (2004). Physics: Frightful, but fun. Pupils' and teachers' views of physics and physics teaching. *Science Education*, 88(5), pp.683-706.
Australian Government (2017). *Australia's National Science Statement*. Canberra: Commonwealth of Australia.
Blair, D. (2018). *Testing the theory: taking Einstein to primary schools*. [online] The Conversation. Available at: https://theconversation.com/testing-the-theory-taking-einstein-to-primary-schools-9710 [Accessed 10 Mar. 2018].
Blair, D., Henriksen, E. and Hendry, M. (2018). *Why don't we teach Einstein's theories in school?*. [online] The Conversation. Available at: https://theconversation.com/why-dont-we-teach-einsteins-theories-in-school-69991 [Accessed 10 Mar. 2018].
Bungum, B., Henriksen, E., Angell, C., Tellefsen, C. and Bøe, M. (2015). ReleQuant – Improving teaching and learning in quantum physics through educational design research. *Nordic Studies in Science Education*, 11(2), p.153.
de Broglie, L. (1970). The reinterpretation of wave mechanics. *Foundations of Physics*, 1(1), pp.5-15.

Choudhary, R., Foppoli, A., Kaur, T., Blair, D., et al. (2018). Can short intervention focused on gravitational waves and quantum physics improve students' understanding and attitude? *Physics Education*

Directorate-General for Education, S. (2018). *Does the EU need more STEM graduates? : final report.*. [online] Publications.europa.eu. Available at: https://publications.europa.eu/en/publication-detail/-/publication/60500ed6-cbd5-11e5-a4b5-01aa75ed71a1/language-en [Accessed 01 Mar. 2018].

Einstein, A. (1905). Über einen die Erzeugung und Verwandlung des Lichtes betreffenden heuristischen Gesichtspunkt. *Annalen der Physik*, 322(6), pp.132-148.

Freidenreich, H., Duncan, R. and Shea, N. (2011). Exploring Middle School Students' Understanding of Three Conceptual Models in Genetics. *International Journal of Science Education*, pp.1-27.

Geelan, D. (1997). Weaving narrative nets to capture school science classrooms. *Research in Science Education*, 27(4), pp.553-563.

Harry, G. (2010). Advanced LIGO: the next generation of gravitational wave detectors. *Classical and Quantum Gravity*, 27(8), p.084006.

Henriksen, E., Bungum, B., Angell, C., Tellefsen, C., Frågåt, T. and Bøe, M. (2014). Relativity, quantum physics and philosophy in the upper secondary curriculum: challenges, opportunities and proposed approaches. *Physics Education*, 49(6), pp.678-684.

Kasen, D., Metzger, B., Barnes, J., Quataert, E. and Ramirez-Ruiz, E. (2017). Origin of the heavy elements in binary neutron-star mergers from a gravitational-wave event. *Nature*.

Kaur, T., Blair, D., Moschilla, J. and Zadnik, M. (2017). Teaching Einsteinian physics at schools: part 2, models and analogies for quantum physics. *Physics Education*, 52(6), p.065013.

Kaur, T., Blair, D., Moschilla, J., Stannard, W. and Zadnik, M. (2017). Teaching Einsteinian physics at schools: part 1, models and analogies for relativity. *Physics Education*, 52(6), p.065012.

Kaur, T., Blair, D., Moschilla, J., Stannard, W. and Zadnik, M. (2017). Teaching Einsteinian physics at schools: part 3, review of research outcomes. *Physics Education*, 52(6), p.065014.

Kelly, L. (2016). *Memory Code*. [S.L.]: Pegasus Books.

Lankford, H. and Wyckoff, J. (1992). Primary and secondary school choice among public and religious alternatives. *Economics of Education Review*, 11(4), pp.317-337.

Lavonen, J., Jauhiainen, J., Koponen, I. and Kurki‐Suonio, K. (2004). Effect of a long‐term in‐service training program on teachers' beliefs about the role of experiments in physics education. *International Journal of Science Education*, 26(3), pp.309-328.

Ligare, M. (2016). Manifestations of classical physics in the quantum evolution of correlated spin states in pulsed NMR experiments. *Concepts in Magnetic Resonance Part A*, 45A(3), p.e21398.

McKagan, S., Perkins, K., Dubson, M., Malley, C., Reid, S., LeMaster, R. and Wieman, C. (2008). Developing and researching PhET simulations for teaching quantum mechanics. *American Journal of Physics*, 76(4), pp.406-417.

Minutephysics. (2012). *Open Letter to the President: Physics Education*. [online] Available at: https://www.youtube.com/watch?v=BGL22PTIOAMM [Accessed 14 Mar. 2018].

Newmann, F. (1975). *Education for citizen action*.

Nobelprize.org. (2018). *The 2017 Nobel Prize in Physics - Press Release*. [online] Available at: https://www.nobelprize.org/nobel_prizes/physics/laureates/2017/press.html [Accessed 01 Mar. 2018].

Pitts, M., Venville, G., Blair, D. and Zadnik, M. (2013). An Exploratory Study to Investigate the Impact of an Enrichment Program on Aspects of Einsteinian Physics on Year 6 Students. *Research in Science Education*, 44(3), pp.363-388.

Stannard, W., Blair, D., Zadnik, M. and Kaur, T. (2016). Why did the apple fall? A new model to explain Einstein's gravity. *European Journal of Physics*, 38(1), p.015603.

Stephenson, J., Strange, V., Allen, E., Copas, A., Johnson, A., Bonell, C., Babiker, A. and Oakley, A. (2008). The Long-Term Effects of a Peer-Led Sex Education Programme (RIPPLE): A Cluster Randomised Trial in Schools in England. *PLoS Medicine*, 5(11), p.e224.

The Conversation. (2010). *The Conversation: In-depth analysis, research, news and ideas from leading academics and researchers.*. [online] Available at: http://theconversation.com/ [Accessed 10 Mar. 2018].

The Einstein-First Project. (2018). *The Einstein-First Project*. [online] Available at: https://www.einsteinianphysics.com/about [Accessed 15 Mar. 2018].

U.S Department of Education (2016). *STEM 2026: A Vision for Innovation in STEM Education*. Washington D.C.

Weule, G. (2017). Gravitational waves and neutron stars: Why this discovery is huge. *ABC News*.

Zahn, C. and Kraus, U. (2014). Sector models—A toolkit for teaching general relativity: I. Curved spaces and spacetimes. *European Journal of Physics*, 35(5), p.055020.